\documentclass[10pt,letterpaper]{article}

\usepackage[margin=1in]{geometry}

\usepackage{amsmath,amssymb}

\usepackage{changepage}

\usepackage{textcomp,marvosym}

\usepackage{nameref,hyperref}

\usepackage[nopatch=eqnum]{microtype}
\DisableLigatures[f]{encoding = *, family = * }

\usepackage[table]{xcolor}

\usepackage{array}
\usepackage{graphicx}
\usepackage{multirow}
\usepackage{csquotes}

\usepackage[numbers,sort&compress]{natbib}

\graphicspath{{Image/}}

\usepackage[labelfont=bf]{caption}

\makeatletter
\renewcommand{\@biblabel}[1]{\quad#1.}
\makeatother

\begin{document}
\vspace*{0.2in}

\begin{flushleft}
{\Large
\textbf\newline{Assessing the informative value of macroeconomic indicators for public health forecasting} 
}
\newline
\\
Shome Chakraborty\textsuperscript{1\Yinyang},
Fardil Khan\textsuperscript{2\Yinyang},
Soutik Ghosal, PhD\textsuperscript{3*}\\
\bigskip
\textbf{1} Gabelli School of Business, Fordham University, New York, New York, United States of America
\\
\textbf{2} Harpur College of Arts and Sciences, Binghamton University, Binghamton, New York, United States of America
\\
\textbf{3} Public Health Sciences Department, School of Medicine, University of Virginia, Charlottesville, Virginia, United States of America
\\
\bigskip

\Yinyang These authors contributed equally to this work.


* soutik.ghosal@virginia.edu

\end{flushleft}
\section{Abstract}
Macroeconomic conditions influence the environments in which health systems operate, yet their value as leading signals of health-system capacity has not been systematically evaluated. In this study, we examined whether certain macroeconomic indicators contained predictive information for several capacity-related public health targets: employment in the health and social assistance workforce, new business applications in the sector, and health care construction spending. Using monthly U.S. time-series data, we evaluated multiple forecasting approaches—including neural network models with different optimization strategies, generalized additive models, random forests, and time series models with exogenous macroeconomic indicators—under different model fitting designs. Across the evaluation settings, we found that macroeconomic indicators provide a consistent and reproducible predictive signal for some public health targets—particularly workforce and infrastructure measures—while other targets exhibit weaker or less stable predictability. Models emphasizing stability and implicit regularization tend to perform more reliably during periods of economic volatility. These findings suggest that macroeconomic indicators may serve as useful upstream signals for digital public health monitoring, while underscoring the need for careful model selection and validation when translating economic trends into health-system forecasting tools.


\section{Introduction}

\subsection{Macroeconomic signals and population health}
The chronic underfunding and fragmentation of public health infrastructure in the United States and globally have left health systems increasingly vulnerable to macroeconomic shocks \citep{ruhm2000recessions, stuckler2009public}. From pandemic-induced supply chain failures to inflation-driven labor shortages, systemic stress often originates far upstream of clinical settings, within economic conditions that destabilize health system operations well before visible breakdowns occur. These disruptions disproportionately affect low-resource settings, where even modest fiscal shocks can cascade into appointment bottlenecks, stockouts of essential drugs, emergency department crowding, and delayed preventive care \citep{miller2020covid}.

Macroeconomic indicators such as retail activity, trade volume, inventory levels, and business formation capture these upstream pressures. Although such indicators are traditionally monitored for fiscal or monetary policy, they also signal downstream constraints on healthcare delivery, workforce stability, and access to care \citep{cawley2003impact}. A contraction in consumer demand, for example, may precede reductions in employer-sponsored insurance coverage or delays in healthcare utilization, while disruptions in trade and logistics can directly affect medical supply chains \citep{cutler2010understanding}. Despite these connections, macroeconomic indicators remain underutilized in population health forecasting frameworks, particularly in systematic, model-based evaluations.

\subsection{Digital data streams for public health forecasting}

The growing availability of high-frequency, digitally collected economic data has created new opportunities for population-level health surveillance. At the same time, advances in machine learning (ML) have expanded the analytical toolkit for modeling complex, nonlinear, and temporally dynamic relationships that characterize both macroeconomic systems and public health targets \citep{varian2014big, mullainathan2017applied, athey2016recursive}.

A substantial literature demonstrates the value of ML methods in economic forecasting. Ensemble models such as random forests and gradient boosting have outperformed traditional linear approaches in applications including GDP nowcasting, inflation forecasting, and macroeconomic risk assessment \citep{richardson2020nowcasting, coulombe2020macroeconomy, yoon2021forecasting}, in part because they capture nonlinear interactions and perform automatic variable selection in high-dimensional settings \citep{chapman2021macroeconomic}.

Related work has extended ML to health-relevant economic and environmental contexts. Satellite imagery has been used to estimate economic activity linked to healthcare access \citep{henderson2012measuring}, computer vision methods have quantified built environments associated with neighborhood health disparities \citep{naik2016cities}, and deep learning applied to central-bank communications has revealed latent economic signals predictive of market behavior \citep{gorodnichenko2023voice}. In public health forecasting, models such as LSTM networks and gradient boosting have improved disease outbreak prediction and hospital resource planning, with reported accuracies exceeding 88\% in some settings \citep{dhanda2025advancement}.

Despite growing recognition of the interdependence between economic conditions and health system capacity—highlighted during shocks such as the COVID-19 pandemic \citep{liu2020methodology}—the integration of structured macroeconomic indicators into public health forecasting remains limited. ML models in this area are increasingly used not as ends in themselves, but as flexible tools for extracting signal from complex data. Prior work has shown that stochastic-gradient–based optimization with regularization and learning-rate control can achieve stable convergence in volatile economic time series \citep{babii2021machine}, while interpretable ensemble and semi-parametric models have supported policy-relevant inference \citep{farbmacher2020explainable, buckmann2021opening}.

In parallel, causal machine learning frameworks—such as causal forests and generalized random forests—have expanded the role of ML in policy analysis \citep{wager2018estimation, athey2019machine}. Although the present study does not seek to estimate causal effects, it aligns with this broader objective by using predictive modeling to inform health-aligned decision-making under uncertainty.

Finally, much of the existing ML literature in public health focuses on clinical prediction or unstructured social determinants (e.g., electronic health records, text, or mobility data). By contrast, structured macroeconomic panel data—while central to fiscal and labor policy—have received comparatively little attention as predictive inputs for public health targets, despite evidence linking economic contractions to healthcare disruptions \citep{stuckler2009public, donaldson2016view, desai2023machine}.

\subsection{Study objectives and contribution}

The goal of this study is not to identify a single optimal predictive model, but to assess how informative macroeconomic indicators are for forecasting public health targets across targets and modeling paradigms \citep{makridakis2020predicting}. For that matter, we consider a variety of ML and statistical models to assess how ranges of macroeconomic indicators can predict the public health targets \citep{athey2019machine}. The list of models include optimization-based neural networks, generalized additive models, time-series models, and ensemble methods—treating these models as complementary analytical lenses. Our primary evaluation is based on a fixed training–testing split, with additional rolling-window strategies used to assess robustness under alternative forecasting scenarios. By comparing error profiles across targets and models, we aim to identify which public health targets exhibit a stable macroeconomic signal. This work contributes to the digital public health literature by systematically evaluating macroeconomic indicators for population health forecasting and assessing robustness across diverse modeling approaches and evaluation mechanisms \citep{soyiri2013overview, bonnet2024scoping}.

\section{Materials and methods}

\subsection{Data description}

\subsubsection{Public health targets}
We examined three targets designed to capture system-level dimensions of public health capacity and access in the United States. These targets were obtained at the monthly level, seasonally adjusted, from the Federal Reserve Bank of St. Louis’s Federal Reserve Economic Data (FRED \citep{fred_unemployment}) platform, which disseminates series produced by U.S. federal statistical agencies, including the U.S. Bureau of Labor Statistics (Current Employment Statistics) and the U.S. Census Bureau. Together, these public health targets represent complementary dimensions of public health system resilience: labor, organizational entry, and physical infrastructure. To ensure a balanced dataset where all targets and features are fully available, we utilized a study period spanning February 2005 through May 2025. Although the balanced dataset is available from July 2004 through May 2025, we begin the usable forecasting sample in February 2005 to keep the fixed-split and rolling-window evaluation periods aligned to consistent calendar boundaries for comparability across models.

\begin{itemize}
    \item \textbf{Health Care and Social Assistance Employees (EM.T)\citep{EMT}}: This variable reflects the operational labor capacity of the U.S. healthcare system. Workforce size is a direct determinant of health service accessibility, provider-to-patient ratios, and surge capacity—especially during public health emergencies. Fluctuations in employment within this sector have been linked to care delays, burnout, and systemic bottlenecks in both urban and rural settings  \citep{buerhaus2017four}.
    
    \item \textbf{Health Care and Social Assistance Business Applications (BA.T)\citep{BAT}}: This indicator captures entrepreneurial activity and new market entries within the healthcare and social assistance sector. It reflects the formation of new clinics, support service providers, and specialty practices, all of which contribute to the system’s flexibility, innovation, and responsiveness to unmet needs. A rise in business applications often precedes greater service availability and diversity, especially in underserved communities \citep{decker2014role}.
    
    \item \textbf{Total Health Care Construction Spending (CTS.T)\citep{CTST}}: Construction spending in the healthcare sector is a key forward-looking indicator of physical infrastructure capacity. It reflects investments in new hospitals, clinics, outpatient centers, and long-term care facilities—facilities critical for expanding care access, reducing geographic disparities, and accommodating future demand. Higher spending often precedes increases in service delivery capability \citep{davis2015despite}.
\end{itemize}

\begin{figure}
    \centering
    \includegraphics[width=0.95\linewidth]{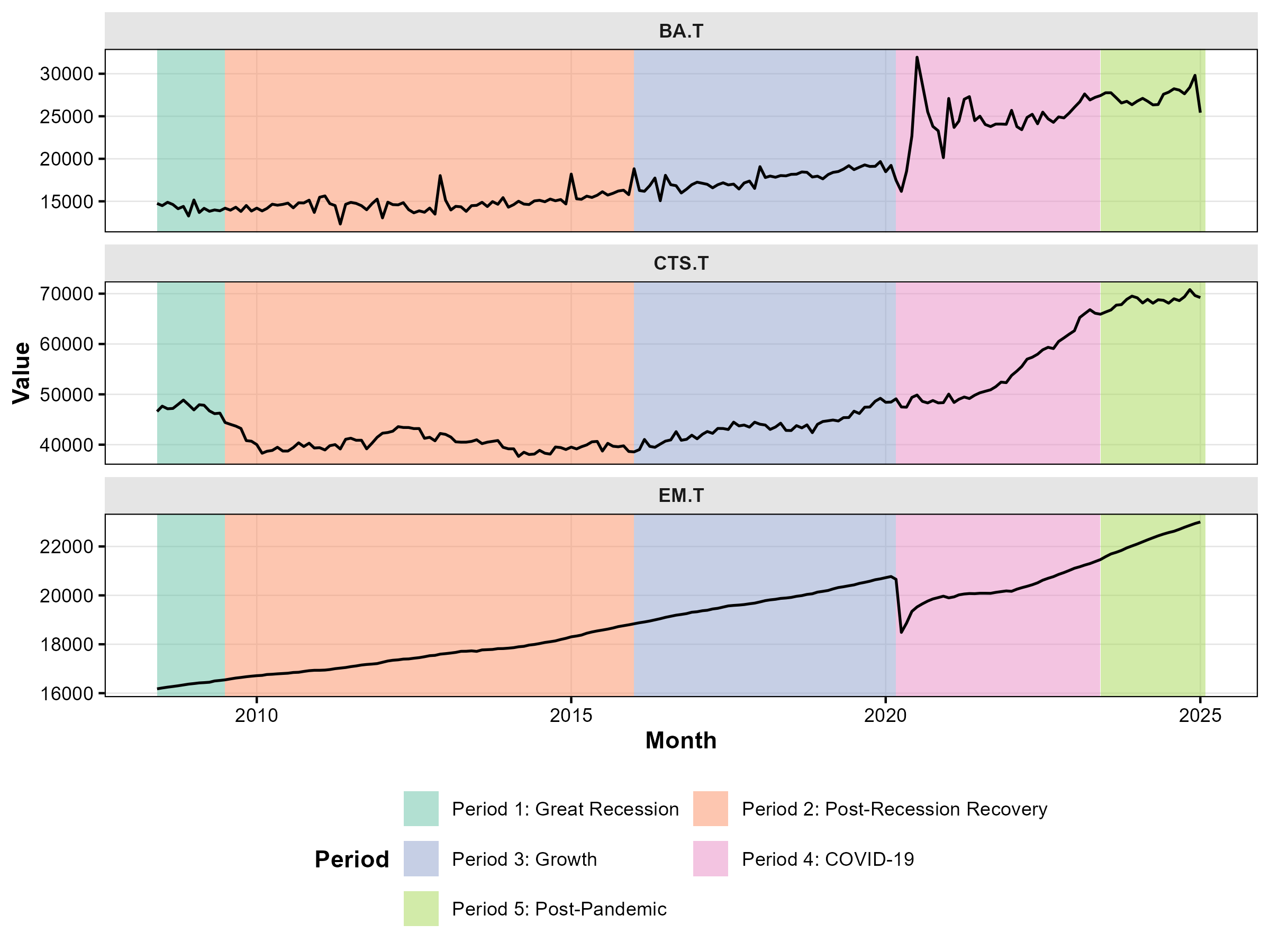}
    \caption{Temporal change of the targets over time periods}
    \label{fig:target_time_series}
\end{figure}

To assess temporal variation in these public health targets, we evaluated their behavior across five U.S. socioeconomic periods. Figure~\ref{fig:target_time_series} depicts their evolution across these periods, defined as follows:
\begin{itemize} 
\item \textbf{Period 1 (June 2008 – June 2009):} The Great Recession 
\item \textbf{Period 2 (July 2009 – December 2015):} Post-Recession Economic Recovery 
\item \textbf{Period 3 (January 2016 – February 2020):} Pre-Pandemic Economic Growth 
\item \textbf{Period 4 (March 2020 – May 2023):} The COVID-19 Pandemic 
\item \textbf{Period 5 (June 2023 – January 2025):} Post-Pandemic Recovery 
\end{itemize}

From Figure~\ref{fig:target_time_series}, we observed that EM.T exhibits a steady upward trajectory from Period 1 through the pre-pandemic period, with a noticeable but temporary decline at the onset of Period 4 (COVID-19), followed by a rapid recovery and continued growth. By contrast, CTS.T declines sharply during Period 1 (Great Recession), then stabilizes with moderate cyclical fluctuations through Periods 2 and 3, before accelerating markedly beginning in the latter half of Period 4. BA.T shows a gradual upward trend through Periods 2 and 3, experiences a brief disruption at the start of Period 4, and then rises sharply during the remainder of the pandemic and into the post-pandemic recovery period.

\subsubsection{Macroeconomic indicators}
The U.S. Census measures 15 indicators of macroeconomic activity, collected from monthly and quarterly economic indicator surveys, available at the Census’s Economic Briefing Room. For each outcome, we use six macroeconomic indicators obtained at the monthly level, seasonally adjusted. Each centers around 1 of 6 core categories of economic activity tracked by the Census (Census Bureau Economic Briefing Room \citep{Census_BriefingRoom}, each relating to a category of economic activity (Business Profits \& Formation, Inventories, Construction \& Housing, Consumer Spending, Manufacturing, and International Trade).

These indicators reflect distinct but interrelated domains of economic activity that plausibly affect public health systems through labor markets, supply chains, consumer demand, and investment patterns. These indicators have been previously linked to upstream determinants of health system capacity and access \citep{donaldson2016view, henderson2012measuring}.

\begin{itemize}
    \item \textbf{Business Applications (BA.F)\citep{BAF}}: Business applications serve as a leading indicator of economic expansion and job creation by capturing entrepreneurial activity and new business formation \citep{decker2014role}.
    \item \textbf{Manufacturing and Trade Inventories (IN.F)\citep{INF}}: Inventory levels signal production adjustments and supply-demand imbalances, serving as an early warning system for economic slowdowns or expansions \citep{uscensus2023mtis}.
    \item \textbf{Total Construction Spending (CTS.F)\citep{CTSF}}: Total construction spending reflects long-term investment in infrastructure and built environment resilience, often preceding employment and service expansion \citep{uscensus2023construction}.
    \item \textbf{Retail and Food Services Sales (CNS.F)\citep{CNSF}}: Retail and food service sales are a key barometer of consumer sentiment and economic momentum, with direct implications for tax revenues and service demand  \citep{uscensus2023retail}.
    \item \textbf{Manufacturers’ Shipments, Inventories, and Orders (MN.F)\citep{MNF}}: These indicators reflect production dynamics and industrial momentum. They are essential for assessing economic cycles and supply chain strength  \citep{uscensus2023m3}.
    \item \textbf{Trade Balance in Goods and Services (IT.F)\citep{ITF}}: The trade balance captures international demand, export competitiveness, and supply chain vulnerability—key factors in healthcare logistics and inflation sensitivity  \citep{bea2023trade}.
\end{itemize}

\begin{figure}
    \centering
    \includegraphics[width=0.95\linewidth]{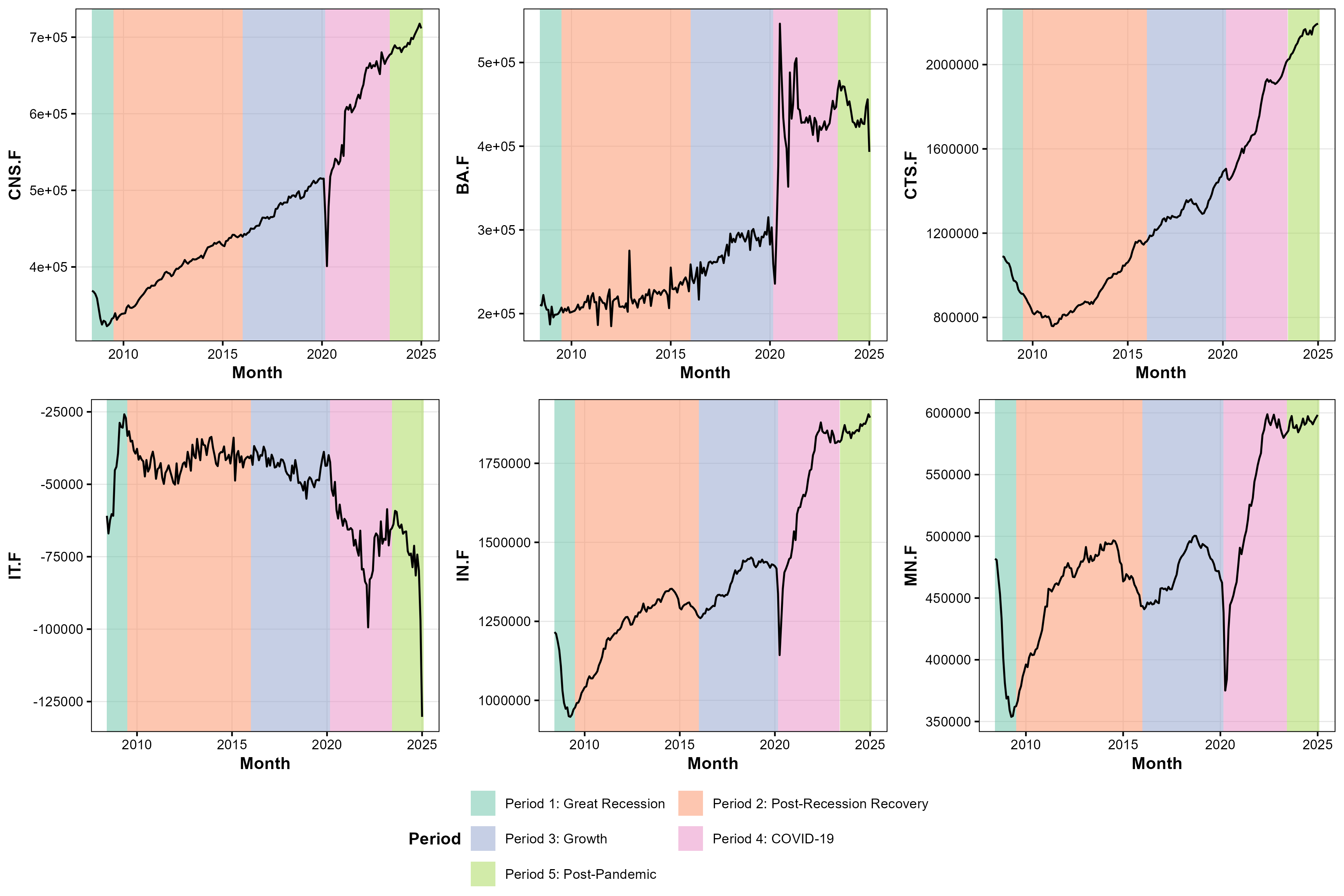}
    \caption{Monthly macroeconomic predictor features (BA.F, IN.F, CTS.F, CNS.F, MN.F, IT.F) over time; shaded regions denote the five socioeconomic periods.}
    \label{fig:feature_time_series}
\end{figure}

Similar to the targets, we have also observed the temporal change of these macroeconomic indicators through Figure~\ref{fig:feature_time_series}. Across the macroeconomic indicators, clear structural shifts are visible across recession, recovery, and pandemic periods. CNS.F shows a steady long-term upward trend from the beginning, with moderate cyclical fluctuations and a sharp but temporary decline at the onset of Period 4 (COVID-19), followed by rapid recovery and continued growth into Period 5. BA.F remains relatively stable through Periods 2 and 3, then exhibits a pronounced spike during the early COVID-19 period, consistent with pandemic-era business restructuring and rapid entry dynamics, before stabilizing at a higher post-pandemic level. CTS.F follows a cyclical investment pattern, declining during Period 1, recovering gradually through Periods 2 and 3, and accelerating sharply through the latter half of Period 4 and into Period 5, reflecting post-pandemic capital expansion. IT.F displays a downward trend through the pre-pandemic period, with pronounced volatility during Periods 3 and 4 and a sharp contraction during COVID-19, consistent with disruptions to global trade flows. IN.F demonstrates long-run growth with cyclical adjustments, interrupted by a brief contraction at the start of Period 4, followed by a steep post-pandemic expansion. Finally, MN.F exhibits repeated expansion–contraction cycles across Periods 2 and 3, a sharp decline at the onset of COVID-19, and strong rebound growth thereafter. Together, these trajectories illustrate coordinated macroeconomic shocks and recoveries that align temporally with major public health and economic crises, reinforcing the plausibility of these indicators as upstream macroeconomic indicators of systemic health capacity.

The detailed summary of the public health targets and the macroeconomic indicators has been tabulated in Table S1 of the supplementary material.



\subsection{Methodology}
\subsubsection{Overview of modeling approaches}

We evaluated multiple statistical and machine learning models to assess whether the predictive value of macroeconomic indicators for public health targets is consistent across different functional forms and learning paradigms \citep{mullainathan2017applied, athey2016recursive, makridakis2020predicting, athey2019machine}. Rather than seeking a single optimal model, this comparative design allows us to identify predictive patterns that persist across model classes. The approaches considered include optimization-based neural networks, generalized additive models, time-series models, and ensemble tree-based methods. Specifically, we consider:

\begin{itemize}
    \item Optimization-based neural network regressors trained using stochastic gradient descent (SGD), Adam, and L-BFGS optimizers \citep{kingma2015adam, liu1989lbfgs};
    \item Generalized additive models (GAMs) to capture nonlinear but interpretable relationships \citep{hastie1986generalized, wood2017generalized}
    \item Autoregressive integrated moving average (ARIMA) models with exogenous regressors \citep{box2015time,hyndman2018forecasting} to account for temporal dependence;
    \item Random forest regression models to capture nonlinear interactions and variable importance \citep{coulombe2020macroeconomy, breiman2001random, liaw2002classification}.
\end{itemize}

\subsubsection{Modeling framework}

\paragraph{Lagged prediction strategy.}
All models are estimated using a common lag structure to ensure the temporal validity of the predictive relationship. For each outcome, macroeconomic indicators observed at month $t-1$ are used to predict the public health outcome measured at month $t$ (or at the beginning of the corresponding forecasting window). This alignment reflects a realistic temporal sequencing in which economic indicators are observed prior to downstream changes in public health targets, and it is consistent with operational forecasting workflows in which economic data become available before health outcomes are measured. All analyses were conducted on monthly time-series panel data consisting of three public health targets and six macroeconomic indicators. Each model uses macroeconomic indicators measured at month $t-1$ to predict the public health target at month $t$, ensuring temporal ordering of predictors and outcomes. It is to be noted that the models used here are for forecasting and association-based interpretation, not causal inference.

\paragraph{Brief mathematical formulation.}
Let $y_t$ denotes a specific public health target at month $t$ and let $\mathbf{X}_{t-1} \in \mathbb{R}^p$ denote the vector of macroeconomic indicators observed at month $t-1$, consistent with the lagged prediction strategy described above. Across all models, the objective is to learn a
mapping
\begin{equation}
y_t = f(\mathbf{X}_{t-1};\,\theta), \label{eqn:main_func}
\end{equation}

where $\theta$ denotes the model parameters, and the prediction error is evaluated using squared or absolute deviations between $y_t$ and $\hat{y}_t$, where $\hat{y}_t$ is the prediction of $y_t$. Below, we provide a brief formulation of each model class; full derivations and implementation details are available in the \href{https://github.com/shomechakraborty/public_health_forecasting_using_computationally_efficient_models.git}{GitHub repository}: \enquote{public\_health\_forecasting\_using\_computationally\_efficient\_models} maintained under the username \enquote{shomechakraborty}.

\paragraph{Neural network models (SGD, Adam, L--BFGS).}

The neural network models assume a nonlinear prediction function for $f(\cdot, \theta)$ in equation \ref{eqn:main_func} as a feed-forward neural network with parameters $\theta$ (weights and biases). The parameters are estimated by minimizing the loss function:
\[
\mathcal{L}(\theta) = \mathcal{L}(y_t, f(\mathbf{X}_{t-1};\,\theta)).
\]
The form of $\mathcal{L}(\theta)$ depends on whether we assume absolute error loss or square error loss. Parameters $\theta$ are updated iteratively using a generic gradient-based rule
\[
\theta_{k+1} = \theta_{k} - \eta_{k}\,\nabla_{\theta}\mathcal{L}(\theta_{k}),
\]
where $\eta_{k}$ denotes the step size (learning rate) and $\nabla_{\theta}\mathcal{L}(\theta_{k})$ is the gradient of training loss.

The three variants of neural network models specified here differ in how the step size and gradient terms are computed. 
\begin{itemize}
\item \textbf{SGD} model updates use the raw gradient with a fixed (or decaying) learning rate. 
\item In the \textbf{Adam} model, the gradient is rescaled using adaptive first- and second-moment estimates, leading to parameter-specific effective learning rates \citep{kingma2015adam}.  
\item In the \textbf{L--BFGS} model, updates follow a quasi-Newton direction that incorporates an approximate curvature (Hessian) adjustment to the gradient, resulting in larger, geometry-aware steps toward the minimizer. 
\end{itemize}

Across all three implementations, the network architecture remains fixed; only the optimization scheme used to obtain $\theta^{*}$ varies.

\paragraph{GAM.}
The GAM specifies a semiparametric additive structure,
\[
y_t
  = \beta_0 + \sum_{j=1}^p s_j(X_{t-1,j}) + \varepsilon_t,
\qquad \varepsilon_t \sim \text{i.i.d. }(0,\sigma^2),
\]
where $\beta_0$ is the intercept, $\epsilon$ corresponds to error, $s_j(\cdot)$ are smooth spline functions estimated via penalized likelihood to balance fit and smoothness.

\paragraph{ARIMA with exogenous macroeconomic indicators.}
The outcome is modeled as a combination of autoregressive dependence, moving-average error dynamics, and lagged macroeconomic predictors:
\[
y_t
 = \sum_{i=1}^p \phi_i\, y_{t-i}
 + \sum_{j=1}^q \psi_j\, \varepsilon_{t-j}
 + \mathbf{X}_{t-1}^{\top}\gamma
 + \varepsilon_t,
\qquad \varepsilon_t \sim \mathcal{N}(0,\sigma^2),
\]
where $\phi_i$ are autoregressive coefficients, $\psi_j$ are moving-average coefficients, and $\gamma$ represents the effects of the lagged macroeconomic indicators $\mathbf{X}_{t-1}$.

\paragraph{Random Forest.}
The Random Forest estimator forms an ensemble prediction
\[
y_t
 = \frac{1}{B}\sum_{b=1}^B f_b(\mathbf{X}_{t-1}),
\]
where each $f_b(\cdot)$ is a regression tree trained on a bootstrap sample with random feature subsampling, allowing the ensemble to capture nonlinear effects and interactions while reducing variance.

Further details of the specified model will be available on GitHub.

\subsubsection{Evaluation design}
We evaluate model performance under two complementary forecasting designs. First, a fixed training–testing split (referred to as 80--20 hereafter), with approximately 80\% of the series used for training the data. It trains models on data from February 2005 to January 2021 and evaluates them on the remaining 20\% of the data, i.e., on February 2021–February 2025. Although the balanced dataset is available from July 2004, we begin the usable forecasting sample in February 2005 to accommodate the one-month lag structure (predictors at $t-1$ used to forecast outcomes at $t$) and to keep the evaluation periods aligned to consistent calendar boundaries.

This setting reflects a standard forward-looking evaluation in which models are trained on historical observations and applied to later periods. Second, a rolling 12–4 design trains on sequential 12-month windows and evaluates on the subsequent non-overlapping 4-month periods. This additional mechanism helps to examine temporal robustness beyond a single static split and to assess the sensitivity of the model performances. We also adapt an additional alternative rolling horizons, such as a 6--1 scheme where models are trained on 6 consecutive months and evaluated on the subsequent 1-month test period. The results from this mechanism are provided in Table S2 and Figure S1 of the Supplement.

\subsubsection{Performance metrics}

The performance of each model for each design mechanism was summarized using mean absolute error (MAE), root mean squared error (RMSE), and N-RMSE (RMSE normalized by the mean of the observed test-period values), which capture the average error magnitude, sensitivity to large deviations, and scale-adjusted error, respectively. For observed public health targets $y_t$ and predictions $\hat{y}_t$ over a test set of size $n$,
\[
\text{MAE}
= \frac{1}{n}\sum_{t=1}^n |y_t-\hat{y}_t|,
\qquad
\text{RMSE}
= \sqrt{\frac{1}{n}\sum_{t=1}^n (y_t-\hat{y}_t)^2 }.
\]
To enable comparison across public health targets with different scales,
\[
\text{N--RMSE}
= \frac{\text{RMSE}}{\bar{y}_{\text{test}}},
\]
where $\bar{y}_{\text{test}}$ is the test-period mean of $y_t$.


\subsubsection{Model implementation}

All models were implemented using widely adopted statistical software to ensure transparency and reproducibility. Default parameter settings were used for baseline estimation, consistent with recommendations to avoid conflating performance differences with extensive hyperparameter tuning \citep{snoek2012practical}. The neural network models (SGD, L--BFGS, and Adam) were implemented in \texttt{Python}, whereas the GAM, ARIMA, and Random Forest models were implemented in \texttt{R}. All analysis scripts and supporting code will be made publicly available in a \href{https://github.com/shomechakraborty/public_health_forecasting_using_computationally_efficient_models.git}{GitHub repository} to facilitate replication and reuse.

\section{Results}

\subsection{Overall predictive performance}

Across the primary 80--20 evaluation, the predictive value of macroeconomic indicators varied across public health targets (Table~\ref{tab:overall_metrics}). Errors were lowest for EM.T across most models, indicating that the macroeconomic indicators used in this study captured a larger share of variation for this target compared with BA.T and CTS.T. By contrast, BA.T exhibited moderately higher error magnitudes, while CTS.T generally showed the largest errors across models, suggesting that this target was more difficult to predict from the available macroeconomic inputs.

Consistent with this pattern, under the 80--20 split the Random Forest achieved an RMSE of 809.41 and N--RMSE of 0.038 for EM.T, compared with an RMSE of 1736.88 and N--RMSE of 0.067 for BA.T. For CTS.T, error magnitudes were substantially larger across models, reflecting weaker alignment between the macroeconomic indicators and this target under the present modeling framework.

No single model uniformly dominated across targets. Rather, performance tended to vary by target (Table~\ref{tab:overall_metrics}), indicating that differences in predictive accuracy were driven more by the underlying predictability of each target than by algorithm choice alone. This pattern is also reflected in Figure~\ref{fig:pred_80_20}, where predicted trajectories for EM.T generally track the observed series more closely than those for BA.T or CTS.T, and greater dispersion across models is visible for the less predictable targets.

Overall, under the 80--20 evaluation, Random Forest and ARIMA were among the most consistent performers for EM.T and BA.T based on RMSE and N--RMSE values, followed by GAM and L--BFGS. For CTS.T under the same scheme, the strongest performance was observed for L--BFGS and Adam, with Random Forest and GAM performing comparably but at higher overall error levels. Under the 12--4 rolling evaluation, the broad ranking of models remained similar, with Random Forest, ARIMA, and L--BFGS generally among the top performers. Results from the 6--1 scheme (reported in the Supplement) showed a similar pattern, although the performance of GAM and ARIMA tended to degrade as the training window shortened, likely reflecting the reduced amount of information available for model fitting. Notably, MAE values were lower in the 12--4 setting than in the 80--20 scheme, suggesting that models trained on shorter, more temporally localized windows sometimes achieved improved predictive performance.

\begin{table}[ht]
\centering
\small
\caption{Overall predictive performance across models and targets under the 80--20 and 12--4 evaluation schemes.}
\label{tab:overall_metrics}
\resizebox{\textwidth}{!}{
\begin{tabular}{clcccccc}
\hline
\multirow{2}{*}{\textbf{Targets}} & \multirow{2}{*}{\textbf{Models}} &
\multicolumn{3}{c}{\textbf{80 -- 20}} &
\multicolumn{3}{c}{\textbf{12 -- 4}} \\
\cline{3-8}
& &
\textbf{MAE (95\% CI)} &
\textbf{RMSE} &
\textbf{N-RMSE} &
\textbf{MAE (95\% CI)} &
\textbf{RMSE} &
\textbf{N-RMSE} \\
\hline

\multirow{6}{*}{\textbf{EM.T}}
 & SGD           & 4459.52 (4328.05, 4590.98) & 4572.99 & 0.215 & 431.40 (389.10, 473.71) & 539.20 & 0.029 \\
 & Adam          & 16242.77 (15835.55, 16649.99) & 16542.75 & 0.777 & 12030.10 (11010.16, 13050.04) & 14336.72 & 0.672 \\
 & L-BFGS        & 3775.77 (3570.46, 3981.09) & 4093.48 & 0.192 & 478.27 (373.44, 583.10) & 933.38 & 0.044 \\
 & GAM           & 1469.19 (1136.64, 1811.67) & 1932.44 & 0.091 & 379.43 (330.69, 437.20) & 627.64 & 0.034 \\
 & ARIMA         & 1321.69 (1050.20, 1598.47) & 1664.03 & 0.078 & 119.05 (82.25, 161.85) & 371.40 & 0.020 \\
 & Random Forest & 722.63 (622.67, 819.71)   & 809.41 & 0.038 & 206.51 (180.32, 236.47) & 284.90 & 0.015 \\
\hline

\multirow{6}{*}{\textbf{BA.T}}
 & SGD           & 4231.46 (3888.93, 4573.98) & 4986.33 & 0.191 & 1552.79 (1245.18, 1860.40) & 2818.38 & 0.156 \\
 & Adam          & 11186.67 (10764.19, 11609.16) & 11650.22 & 0.447 & 12968.64 (12065.08, 13872.20) & 14694.10 & 0.562 \\
 & L-BFGS        & 2376.28 (2111.66, 2640.90) & 3130.45 & 0.120 & 1237.11 (1005.69, 1468.54) & 2159.09 & 0.083 \\
 & GAM           & 3819.55 (3419.42, 4245.53) & 4113.80 & 0.158 & 3118.54 (2670.10, 3606.71) & 5323.15 & 0.292 \\
 & ARIMA         & 1950.86 (1582.98, 2337.85) & 2403.03 & 0.092 & 1712.03 (1416.75, 1979.54) & 2815.68 & 0.154 \\
 & Random Forest & 1547.64 (1332.18, 1763.15) & 1736.88 & 0.067 & 873.88 (706.23, 1068.22) & 1685.79 & 0.093 \\
\hline

\multirow{6}{*}{\textbf{CTS.T}}
 & SGD           & 18905.24 (17632.50, 20177.99) & 21295.07 & 0.346 & 2742.23 (2447.00, 3037.45) & 3551.82 & 0.076 \\
 & Adam          & 4184.56 (3867.89, 4501.22) & 4843.33 & 0.079 & 2050.68 (1838.15, 2263.21) & 2616.49 & 0.042 \\
 & L-BFGS        & 3140.23 (2824.58, 3455.89) & 3971.19 & 0.065 & 1539.46 (1321.56, 1757.37) & 2268.47 & 0.037 \\
 & GAM           & 8345.71 (6750.99, 10000.09) & 10156.23 & 0.165 & 2362.32 (2090.28, 2643.91) & 3224.32 & 0.069 \\
 & ARIMA         & 13163.08 (11107.35, 15248.73) & 15133.57 & 0.246 & 1864.03 (1648.46, 2113.62) & 2637.73 & 0.056 \\
 & Random Forest & 8055.80 (6718.66, 9435.21) & 9447.69 & 0.153 & 1360.45 (1199.82, 1516.71) & 1813.79 & 0.039 \\
\hline
\end{tabular}
}
\end{table}

\subsection{Temporal robustness and prediction trajectories}

\begin{figure}
    \centering
    \includegraphics[width=0.95\linewidth]{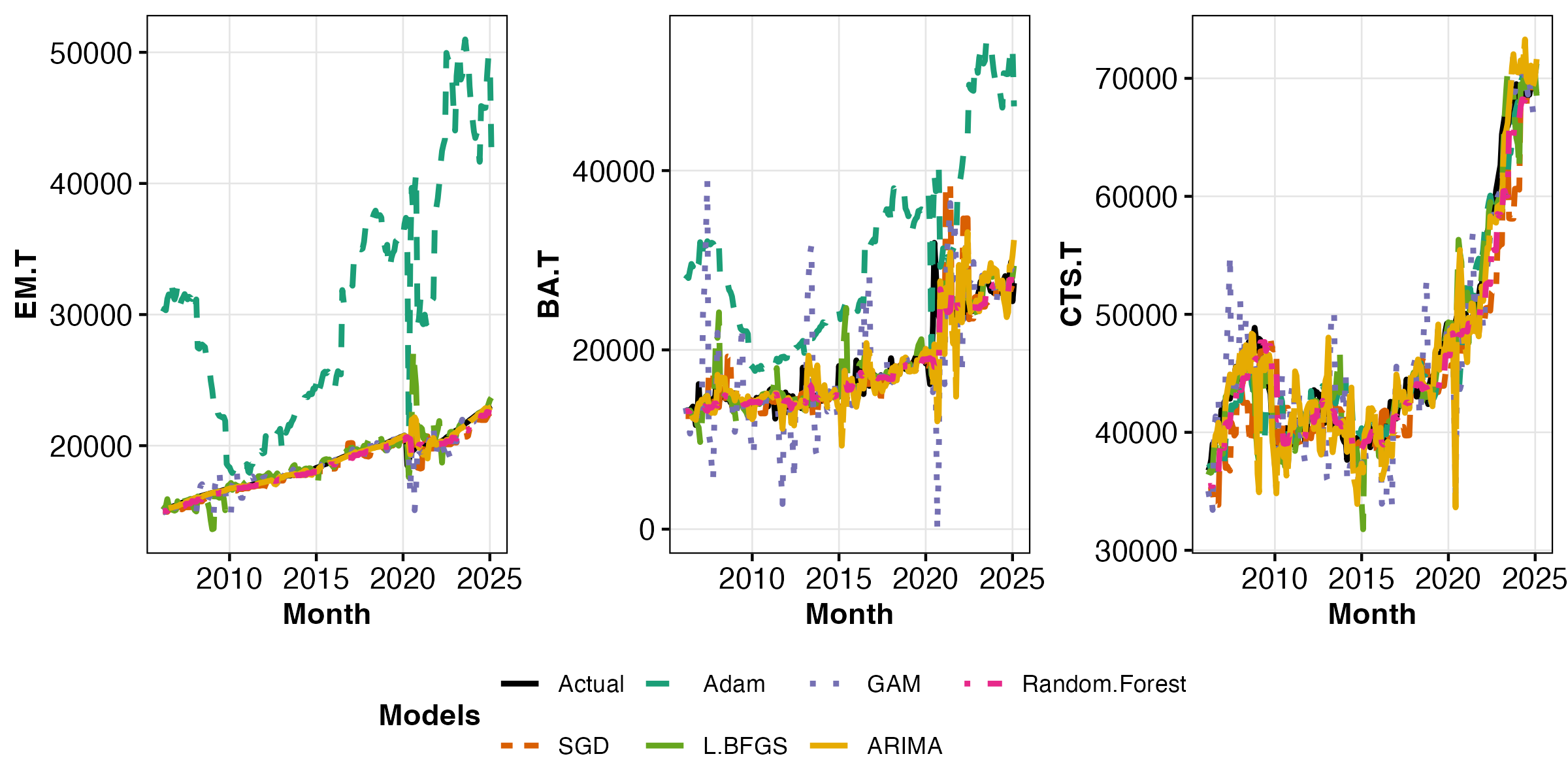}
    \caption{Comparing predicted targets from different models under the 12--4 mechanism}
    \label{fig:pred_12_4}
\end{figure}

Findings were broadly consistent under the rolling 12--4 evaluation (Figure~\ref{fig:pred_12_4}). Public health targets that showed strong macroeconomic signal in the 80--20 analysis (e.g., EM.T and CTS.T) continued to produce stable prediction accuracy across rolling windows, whereas BA.T displayed greater temporal variability and smoother but less responsive prediction trajectories. As expected, absolute error levels increased modestly due to shorter training horizons, but the relative ordering of target difficulty and model performance remained similar to the primary evaluation.

\begin{figure}
    \centering
    \includegraphics[width=0.95\linewidth]{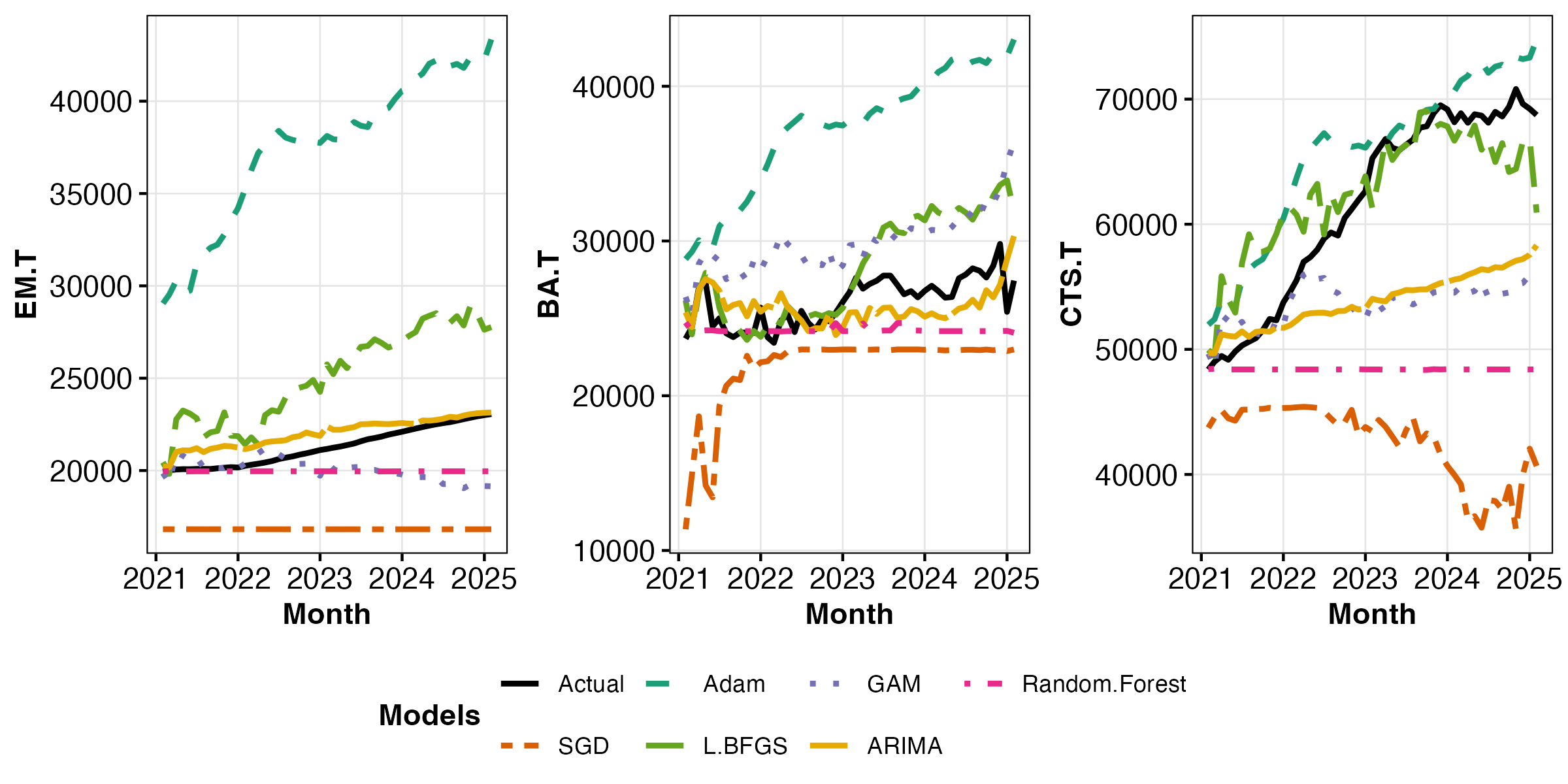}
    \caption{Comparing predicted targets from different models under the 80--20 mechanism}
    \label{fig:pred_80_20}
\end{figure}

Taken together, these results indicate that macroeconomic indicators contain meaningful and reproducible predictive information for certain health system targets—particularly those related to workforce and infrastructure capacity—while predictive value is more limited for others. The stability of findings across models and evaluation strategies supports the interpretation that performance patterns reflect underlying macroeconomic signals rather than model-specific artifacts. These findings support the interpretation that observed predictive performance (Figure~\ref{fig:pred_80_20}) reflects intrinsic signal strength rather than overfitting or algorithmic artifacts.

\subsection{Results from other mechanism}

Under the 6–-1 rolling evaluation scheme—where models are trained on six months of data and tested on the subsequent month—performance patterns were broadly consistent with those observed in the primary analyses. As summarized in Table S2 and illustrated in Figures S1–S2, the Random Forest model demonstrated the strongest overall performance across targets, yielding the lowest MAE and RMSE values for EM.T, BA.T, and CTS.T (e.g., EM.T MAE = 111.94; N-RMSE = 0.010). The L-BFGS neural model also performed competitively for CTS.T and EM.T, whereas Adam and GAM consistently exhibited poorer accuracy, with substantially higher errors across public health targets. These 6–1 results, included in the Supplementary Materials, reinforce the robustness of the core findings across alternative temporal evaluation frameworks.



\section{Discussion}

Our results indicate that some public health targets exhibit stronger and more stable predictive alignment with macroeconomic indicators than others, particularly under forecasting-oriented evaluation designs. These relationships are most evident for targets linked to structural system conditions such as labor capacity, market activity, and infrastructure investment. The findings should be interpreted as evidence of predictive associations rather than causal effects. The models evaluate whether macroeconomic indicators contain signals useful for forecasting public health targets, rather than identifying economic mechanisms or intervention effects. Across both evaluation schemes (the 80--20 fixed split and the rolling 12--4 framework), models demonstrated consistent directional alignment between macroeconomic features and health system targets, while also revealing periods in which performance differed in response to volatility and structural regime shifts. These results reinforce evidence that economic fluctuations propagate through workforce, capital, and service-delivery channels in ways that shape downstream public health access and resilience \citep{coulombe2020macroeconomy, babii2021machine, russo2021impact, cawley2003impact,kruk2015resilient}.

A central contribution of this work is the comparative evaluation of forecasting models in a health-relevant macroeconomic context. More flexible learning architectures (such as Adam-optimized neural networks and GAM-based models) were able to capture nonlinear temporal variation, but in some settings their performance degraded during shock and transition periods, particularly around the COVID-19 disruption. This observation is consistent with prior work suggesting that aggressively adaptive
optimization strategies may be less stable in nonstationary environments \citep{wilson2017marginal, keskar2016large}. By contrast, models with stronger implicit regularization and smoother optimization dynamics, including SGD-based and L--BFGS neural models, demonstrated more stable error behavior and closer alignment with observed outcome trajectories across targets, echoing prior findings on the generalization advantages of constrained or structure-aware training in economic time-series modeling.

These differences are not only methodological but also practically relevant for health-system planning. Forecasting tools intended to support preparedness and resource allocation must balance predictive accuracy with reliability, computational tractability, and transparency—particularly in settings with limited analytic capacity. Models that perform consistently across economic regimes are better positioned to inform anticipatory decisions such as workforce stabilization, infrastructure investment timing, and surge-capacity planning, aligning with broader priorities in digital health regarding deployability and equity-sensitive decision support.

investment rebounds, and pandemic-era reallocations in capital and service capacity. These concordances increase confidence that the models are capturing structural linkages rather than spurious statistical associations.


Several design choices introduce important limitations. First, the analysis uses a one-month lag structure; alternative lag horizons or distributed-lag specifications may reveal different temporal relationships. Second, the set of macroeconomic indicators and public health targets is necessarily selective, and different indicator combinations may yield different forecasting values. Third, model evaluation focuses on aggregate predictive accuracy rather than feature-specific attribution, and future work should examine which indicators contribute most strongly to forecast performance across periods. Fourth, the findings are strictly based on the economy and public health infrastructure within the US and, hence, may not generalize to other countries or health-system contexts. Finally, the performance of classical statistical models was sensitive to the evaluation design. Under the rolling 12--4 scheme, ARIMA produced missing predictions in early test windows due to standard initialization constraints when fitting ARIMA models on short training sequences, an issue not observed under the fixed 80--20 split, where longer historical series were available. GAM performance similarly degraded under shorter training horizons (e.g., the 6--1 scheme), reflecting limited information for stable smooth estimation. Together, these patterns highlight the greater robustness of the machine learning models across varying training--testing configurations. Future work should explore alternative lag structures, cross-context generalization, and deployment-focused evaluation to assess whether such forecasting pipelines can be integrated into operational public-health monitoring environments.


From a digital-health perspective, these results suggest that routinely collected macroeconomic indicators may serve as an upstream data stream for monitoring structural stress in health-system capacity \citep{McIntyre2023}. Rather than replacing clinical surveillance, such forecasts may complement operational planning by identifying periods when capacity trends begin to diverge from historical patterns.

\section{Conclusion}

This study shows that macroeconomic indicators can provide useful lead-time signals about structural conditions in the U.S. health system and that these signals can be translated into meaningful forecasts using computationally efficient and statistically stable modeling approaches. Across two complementary evaluation frameworks and multiple model classes, we observed consistent
predictive alignment between economic precursors and health-system targets, with models emphasizing stability and implicit regularization performing more reliably during periods of volatility.

The contribution of this work lies not only in model comparison but in framing macroeconomic forecasting as a component of digital public health infrastructure. The results suggest that economic surveillance can complement clinical and epidemiologic monitoring by supporting anticipatory awareness of system-level pressures relevant to planning, preparedness, and policy response. At the same time, the findings should be interpreted in light of key constraints, including the reliance on a single-lag specification, a selected set of macroeconomic indicators and capacity-focused outcome measures, and the absence of causal identification or feature-level attribution. These factors highlight the need for cautious interpretation and motivate extensions that explore broader indicator sets, alternative lag structures, and outcome domains.

Future research should build on this work by integrating macroeconomic signals with other data streams, examining feature- and mechanism-specific pathways, and evaluating transportability across institutional and geographic contexts. Such efforts will help advance economic-to-health forecasting as a practical, equity-aware digital health capability that supports more anticipatory and resilient health-system governance.


\bibliographystyle{plos2015}
\bibliography{references}  

\end{document}


\maketitle

\section*{Input and Target Data Summary.}
This section provides a summary of the minimum, maximum, mean, and standard deviations of each input and target between February, 2005 and February, 2025. The summaries are tabulated in Table~\ref{tab:Summaries}.
\begin{table}[!ht]
\begin{center}
\resizebox{0.95\textwidth}{!}{%
    \begin{tabular}{ccccccc}
\hline
\textbf{Variable}      & \textbf{Statistic} & \textbf{Period 1} & \textbf{Period 2} & \textbf{Period 3} & \textbf{Period 4} & \textbf{Period 5} \\ \hline
\multirow{4}{*}{EM.T}  & Max                & 16520.30          & 18795.60          & 20771.00          & 21381.00          & 23001.20          \\
                       & Mean               & 16355.77          & 17528.81          & 19773.45          & 20254.67          & 22282.10          \\
                       & Min                & 16176.70          & 16544.80          & 18838.10          & 18485.00          & 21463.10          \\
                       & SD                 & 108.66            & 628.80            & 548.90            & 629.72            & 469.64            \\ \hline
\multirow{4}{*}{BA.T}  & Max                & 15156.00          & 18204.00          & 19674.00          & 31945.00          & 29812.00          \\
                       & Mean               & 14250.38          & 14765.90          & 17733.10          & 24578.03          & 27308.70          \\
                       & Min                & 13267.00          & 12326.00          & 15067.00          & 16166.00          & 25423.00          \\
                       & SD                 & 531.35            & 894.26            & 1029.60           & 2867.92           & 961.54            \\ \hline
\multirow{4}{*}{CTS.T} & Max                & 48862.00          & 44406.00          & 49216.00          & 66814.00          & 70810.00          \\
                       & Mean               & 47321.62          & 40376.58          & 43552.58          & 54195.31          & 68475.20          \\
                       & Min                & 46162.00          & 37660.00          & 38568.00          & 47446.00          & 65920.00          \\
                       & SD                 & 792.07            & 1592.73           & 2512.30           & 6091.97           & 1165.32           \\ \hline
\multirow{4}{*}{BA.F}  & Max                & 222233.00         & 275276.00         & 315189.00         & 546415.00         & 478209.00         \\
                       & Mean               & 204100.62         & 217722.45         & 275508.88         & 423143.33         & 443790.15         \\
                       & Min                & 187029.00         & 184968.00         & 216685.00         & 235695.00         & 393232.00         \\
                       & SD                 & 8679.08           & 13791.56          & 20851.16          & 57686.82          & 21902.66          \\ \hline
\multirow{4}{*}{IN.F}  & Max                & 1215940.00        & 1353047.00        & 1452088.00        & 1879425.00        & 1905140.00        \\
                       & Mean               & 1052088.15        & 1227082.67        & 1377235.30        & 1653458.87        & 1858472.90        \\
                       & Min                & 947862.00         & 977575.00         & 1260021.00        & 1143208.00        & 1817501.00        \\
                       & SD                 & 107134.51         & 105696.56         & 64361.73          & 199812.76         & 22756.10          \\ \hline
\multirow{4}{*}{CTS.F} & Max                & 1090098.00        & 1164377.00        & 1498345.00        & 2011831.00        & 2194337.00        \\
                       & Mean               & 1005574.39        & 915880.28         & 1312170.40        & 1731137.97        & 2123890.00        \\
                       & Min                & 912530.00         & 758376.00         & 1160988.00        & 1452280.00        & 2023013.00        \\
                       & SD                 & 64181.96          & 119497.04         & 82696.31          & 184285.86         & 54646.65          \\ \hline
\multirow{4}{*}{CNS.F} & Max                & 368926.00         & 442149.00         & 515866.00         & 680253.00         & 717662.00         \\
                       & Mean               & 340606.31         & 393651.97         & 479771.28         & 603169.46         & 693016.40         \\
                       & Min                & 322634.00         & 330998.00         & 439466.00         & 401028.00         & 677117.00         \\
                       & SD                 & 18031.48          & 33702.61          & 23247.48          & 67916.56          & 11843.89          \\ \hline
\multirow{4}{*}{MN.F}  & Max                & 481867.00         & 496723.00         & 500424.00         & 598859.00         & 598328.00         \\
                       & Mean               & 405165.54         & 456423.22         & 471176.10         & 527132.67         & 591411.80         \\
                       & Min                & 353787.00         & 362449.00         & 440948.00         & 375078.00         & 583700.00         \\
                       & SD                 & 50700.98          & 35451.22          & 19306.58          & 65271.03          & 4401.67           \\ \hline
\multirow{4}{*}{IT.F}  & Max                & -25840.00         & -31674.00         & -36801.00         & -42374.00         & -59165.00         \\
                       & Mean               & -44794.00         & -41257.72         & -44333.16         & -68129.90         & -73724.95         \\
                       & Min                & -66989.00         & -50126.00         & -54997.00         & -99398.00         & -130273.00        \\
                       & SD                 & 15642.33          & 4304.58           & 4294.84           & 10820.17          & 16006.86          \\ \hline
\end{tabular}}
    \caption{Summaries of targets and features across periods.}
    \label{tab:Summaries}
    \end{center}
\end{table}




\section*{6-1 Model Performance Summary.}
This section provides a summary of the model error metrics on each target. From Table~\ref{tab:overall_metrics_6_1}, we observe that the Random Forest model performed the best across all targets, with N-RMSEs of around 0.010, 0.078, and 0.024 for EM.T, BA.T, and CST.T. Based on the optimal N-RMSEs of the random forest model in each target, the inputs carry the strongest predictive signaling for CST.T, followed by BA.T and then EM.T. The ARIMA model was not part of this analysis because of short training sequences. Furthermore, the GAM also showed extremely poor behavior, unlike the 80 -- 20 mechanism, where the performance of GAM was adequate. These collectively show how a very short training window impacts the statistical models.

\begin{table}[ht]
\centering
\small
\caption{Overall predictive performance across models and targets under the 6--1 evaluation schemes.}
\label{tab:overall_metrics_6_1}
\resizebox{\textwidth}{!}{
\begin{tabular}{ccccc}
\hline
\multirow{2}{*}{\textbf{Targets}} & \multirow{2}{*}{\textbf{Metrics}} & \multicolumn{3}{c}{\textbf{6 -- 1}}                             \\ \cline{3-5} 
                                  &                                   & \textbf{MAE (95\% CI)}        & \textbf{RMSE} & \textbf{N-RMSE} \\ \hline
\multirow{5}{*}{\textbf{EMT}}     & SGD                               & 196.65 (167.19, 226.11)       & 301.67        & 0.016           \\
                                  & Adam                              & 12581.96 (11567.98, 13595.94) & 14842.12      & 0.801           \\
                                  & L-BFGS                            & 244.7 (200.57, 288.82)        & 421.034       & 0.023           \\
                                  & GAM                               & 1079.68 (761.31, 1474.97)     & 3247.867      & 0.175           \\
                                  & Random Forest                     & 111.94 (97.27, 131.24)        & 186.541       & 0.010           \\ \hline
\multirow{5}{*}{\textbf{BAT}}     & SGD                               & 1229.67 (977.37, 1481.97)     & 2312.91       & 0.129           \\
                                  & Adam                              & 13518.6 (12610.54, 14426.65)  & 15246.71      & 0.850           \\
                                  & L-BFGS                            & 1277.44 (968.81, 1586.06)     & 2715.528      & 0.151           \\
                                  & GAM                               & 6189.87 (4977.66, 7990.68)    & 13814.56      & 0.764           \\
                                  & Random Forest                     & 762.61 (626.69, 914.04)       & 1402.622      & 0.078           \\ \hline
\multirow{5}{*}{\textbf{CTST}}    & SGD                               & 1468.84 (1309, 1628.68)       & 1922.935      & 0.042           \\
                                  & Adam                              & 1222.21 (1091.51, 1352.91)    & 1588.595      & 0.034           \\
                                  & L-BFGS                            & 1096.82 (947.52, 1246.13)     & 1595.913      & 0.034           \\
                                  & GAM                               & 9591.02 (6319.95, 14419.29)   & 35309.43      & 0.757           \\
                                  & Random Forest                     & 869.14 (785.83, 962.58)       & 1102.694      & 0.024           \\ \hline
\end{tabular}
}
\end{table}

\begin{figure}
    \centering
    \includegraphics[width=0.95\linewidth]{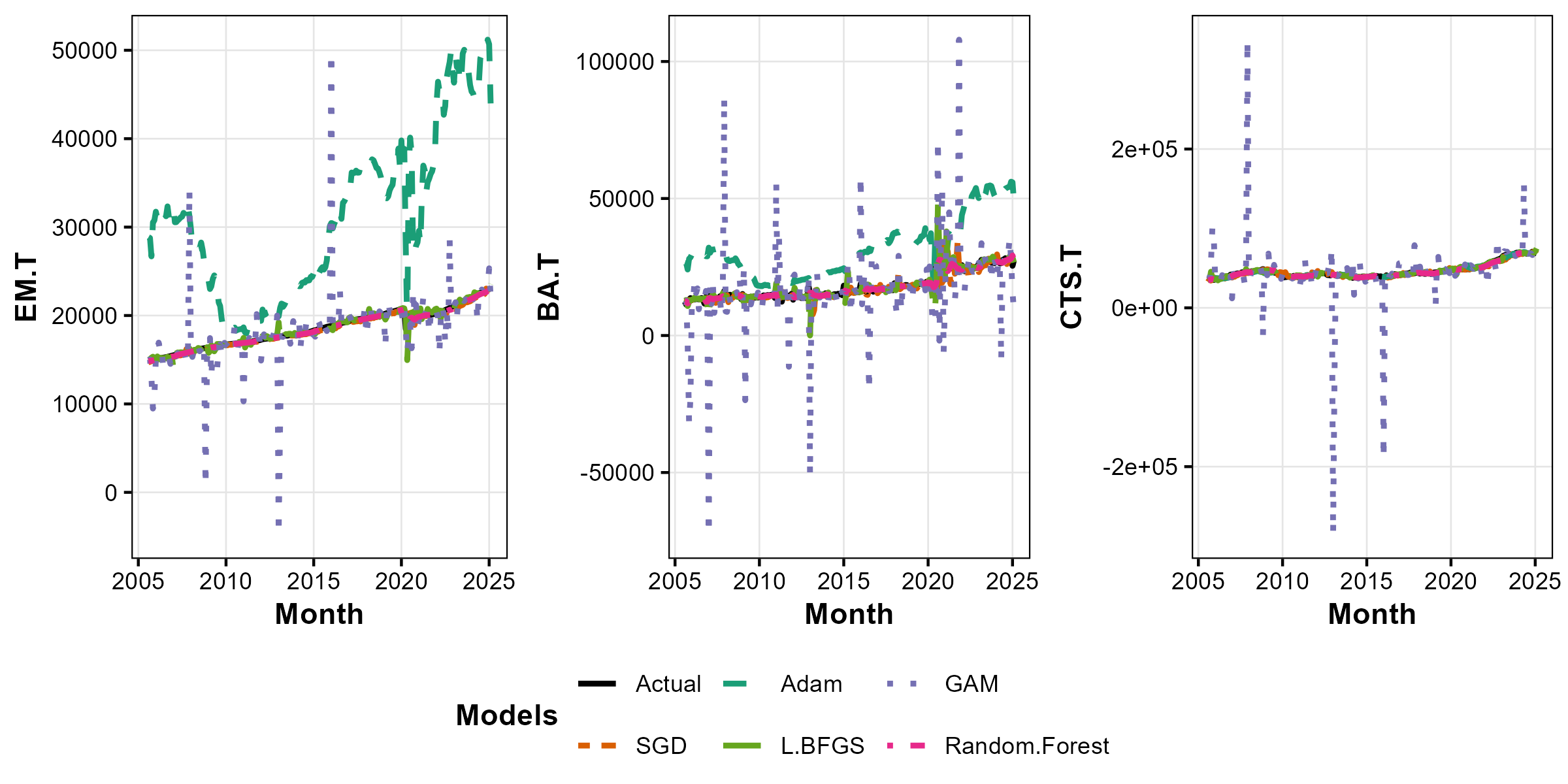}
    \caption{Comparing predicted targets from different models under the 6--1 mechanism}
    \label{fig:pred_6_1}
\end{figure}

\begin{figure}
    \centering
    \includegraphics[width=0.95\linewidth]{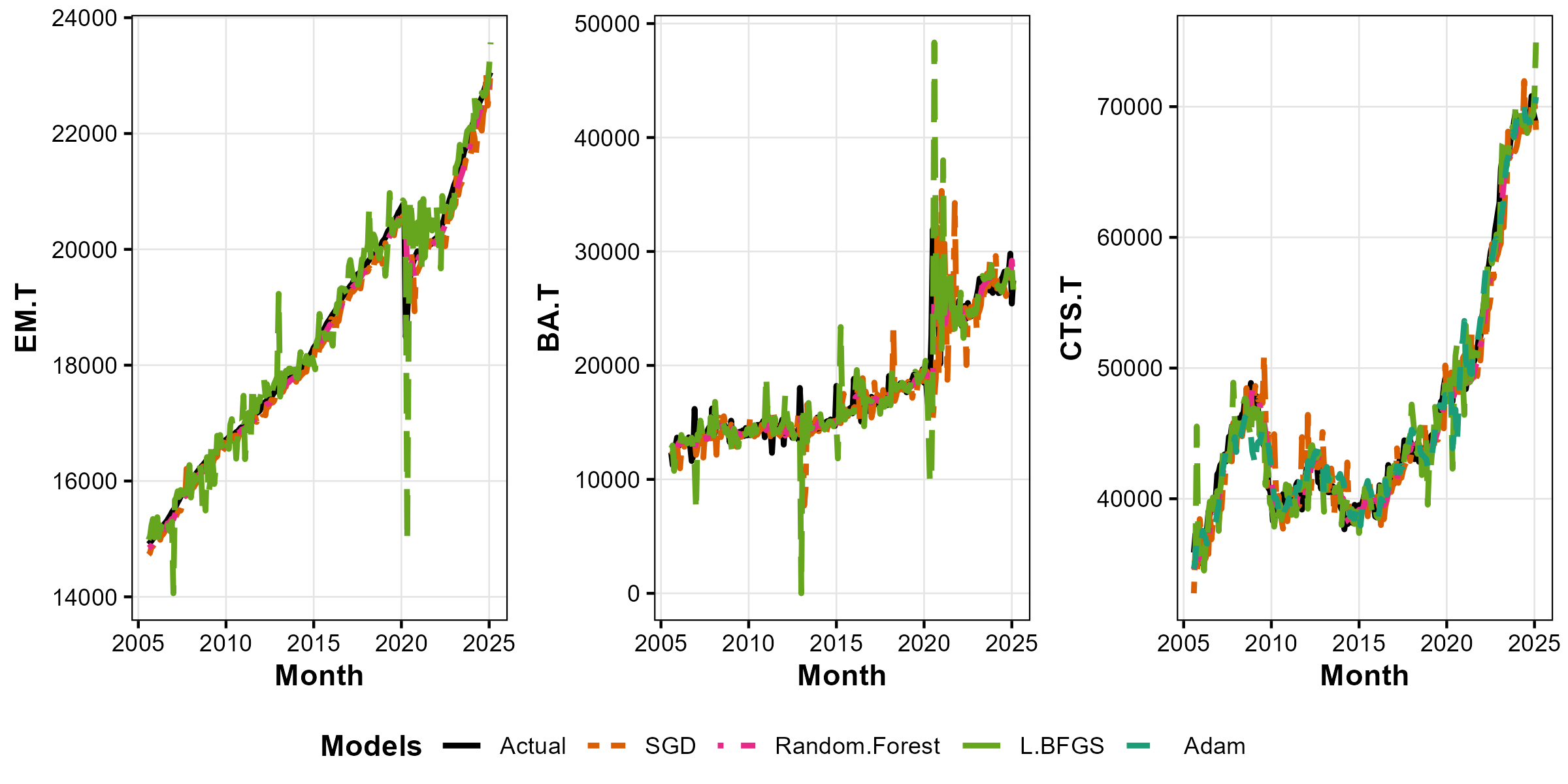}
    \caption{Comparing predicted targets from best performing models under the 6--1 mechanism}
    \label{fig:pred_6_1_best}
\end{figure}

We also plot the predicted target values across different models for the  -- 1 mechanism. Figure~\ref{fig:pred_6_1} shows the performances of all the models, whereas Figure~\ref{fig:pred_6_1_best} only highlights the top-performing ones for each of the targets.

